\documentclass[a4paper,11pt]{article}
\usepackage[utf8]{inputenc}
\usepackage[margin=0.95in]{geometry}
\usepackage{graphicx}
\graphicspath{figures/} 
\usepackage[font={rm,md,it}]{subfig}
\usepackage[authoryear,round,longnamesfirst]{natbib}
\usepackage{booktabs}

\title{Effects of Visual Priming on Rating Scale Usage}
\author{Pieter C. Schoonees \and Patrick J.F. Groenen \and Michel van de Velden \and Hester van Herk}

\begin{document}

\maketitle

\begin{abstract}
    Rating scales are much used in survey research. Often, it is assumed that the scores obtained through rating scales can be compared within and between respondents when studies are in one country. In addition, it is assumed that they can be treated as a numerical scale. In this paper, we study the anchoring effect of a visual stimulus on rating scale usage. To do so, we set up a randomized experiment where the experimental group was primed by asking to rate the filling of a cylinder that was presented visually. For a five point rating scale, we find the effect that primed respondents use Category 1 less and Categories 3 and 4 more, and no effect on Categories 2 and 5. 
\end{abstract}

\section{Introduction}

The analysis and in particular the quality of survey data is an important topic in social science research to which J\"{o}rg Blasius has made many contributions over the years \citep[see, for example,][]{blasius2012}. Here, too, we study the possibilities for improving quality of respondents' ratings. In particular, we investigate one of the key assumptions of most analyses of surveys using rating scales is that the  answers on the items can be compared within and between respondents. Often these assumptions are not tested when comparing respondents in the same country. One source of differences in rating scale usage by respondents is due to various response styles, such as extreme scoring, acquiescence, and midpoint scoring \citep{herk2004}. 
Several solutions have been proposed for dealing with response styles. A simple one is ipsatizing \citep{cunningham1977,rudnev2021}. 

Technical solutions to response style problems include searching for a priori unknown response style patterns through the latent class bilinear multinomial logit model \citep{vanRosmalen2010identifying} and the constrained dual scaling model by \citet{schoonees2015constrained}. In this paper, we study a different approach, that of priming the respondents by a visual task that aims at anchoring the rating scale to a clear visual stimulus. The aim is to use priming for calibrating respondents to the same scale usage. To do so, we investigate through an experiment whether there is an effect of anchoring, here, priming the respondents with a task where they need to rate the amount of filling of a cylinder by indicating how much of the cylinder was filled with red (see Figure~\ref{F:cylinder} for an example). After this task, the questionnaire continued with questions using 5-point rating scales.

In the remainder of this paper, we discuss first how the data were gathered, the experiment was set up, and how the priming through a visual stimulus (the amount of filling of a cylinder) was done. Then, the consistency of the answers in the experimental condition (rating versus percentage) is explored. After that, we perform two tests to compare the distributions over the average rating scale categories between the treatment and control groups. Subsequently, the difference in distributions is explored at the item level to see, amongst others, whether there is a diminishing effect over time. Finally, a correspondence analysis is used to study additional item differential effects. We end with a discussion and conclusions.

\section{Materials and Methods}
The data were collected in 2013 by GfK Panel Services Benelux B.V., a professional marketing research company. 
The total number of respondents in this research is 1621. They were 58.5\% females, 13.6\% 18-29 years, 17.2\% 30-39 years, 20.4\% 40-49 years, 27.1\% 50-64 years, and 21.7\% 65 years or older, meaning that they are a bit older than the general Dutch population. Of these respondents, roughly half (786) were presented with the cylinder questions (the treatment group). The other 835 respondents did not receive these questions (control group). Individuals were randomly allocated to one of the groups, with no statistically significant differences in gender, age group, or education level. 



\begin{figure}
\centering
    \includegraphics[height=0.35\textwidth]{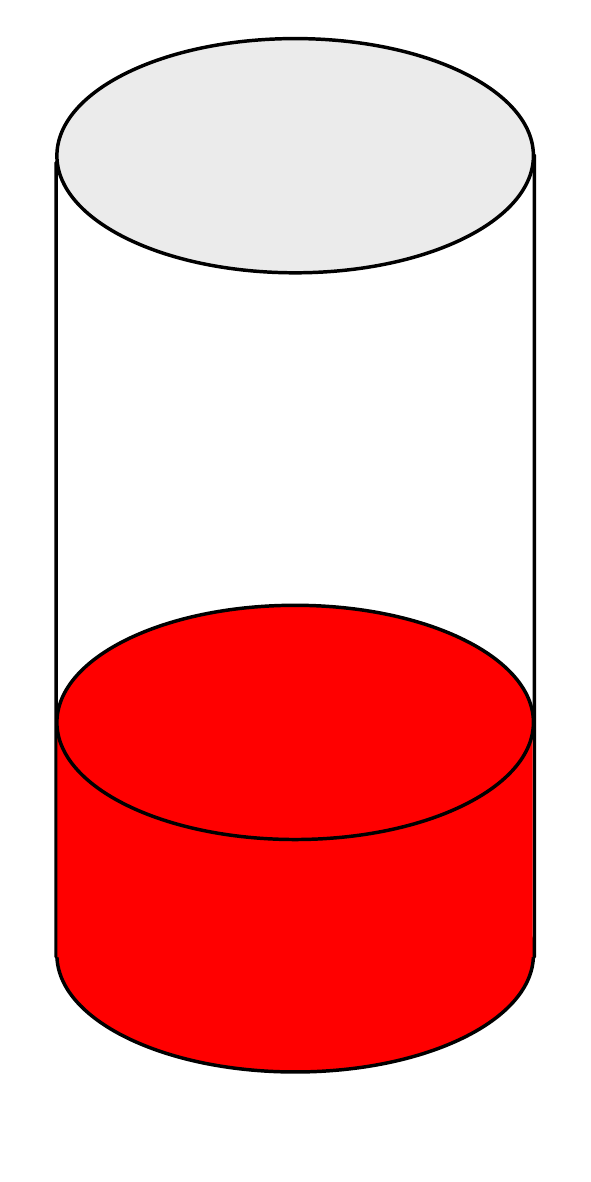}
\caption{Example of visual stimulus of a 30\% red-filled cylinder used in one of the treatment items. The respondent was asked to estimate the amount of red in the cylinder on a five-point rating scale.}
\label{F:cylinder}
\end{figure}

In the questionnaire, the first five survey items concerned the respondent's economic outlook for the Dutch economy, which was measured on a three-point scale with categories ``better", ``worse," or ``unchanged". These items were answered by all respondents. For the treatment group, this was followed by the 10 cylinder questions (see Figure~\ref{F:cylinder} for an example), which consisted of two blocks of five questions each. In the first block, respondents were to assess the amount of red on each cylinder on a 5-point scale ranging from ``almost no red" (1) to ``almost entirely red" (5), with the middle categories not labelled. The true percentages of the amount of red in the cylinders was 10\%, 30\%, 50\%, 70\%, and 90\% corresponding naturally to a five point rating scale. In the second block of five questions, the respondents were asked to give percentage (0--100) to indicate the redness of the cylinder. The questions within each of these two blocks were presented in random order, and each cylinder was displayed for one second. 
The control group did not receive the 10 cylinder items. After the cylinders, both groups were asked to assess 16 items from the marketing literature on innovativeness \citep{steenkamp1999} and materialism \citep{richins1992}. These multi-item scales were measured on a 5-point Likert scale (ranging from ``totally agree" to ``totally disagree"). Finally, all respondents were asked to answer 4 questions on sustainable behavior by ticking one of the options ``never", ``sometimes", ``often", ``very often" or ``almost always". 

In the next sections, we mainly focus on the 16 items of the innovativeness and materialism scales as these items followed directly on the cylinder items in the treatment group and used the same five point rating scale.

\section{Results}
First, we study the treatment group in their consistency of the rating scale usage and their estimated percentages of red. Then we focus whether or not there is a difference in the distribution in rating scale usage on the 16 items between treatment and control groups. After that, we study the differences on an item level.

\subsection{Answer Consistency within the Treatment Group}
\label{S:consistency}

\begin{figure}
\centering
 \includegraphics[width=0.6\textwidth]{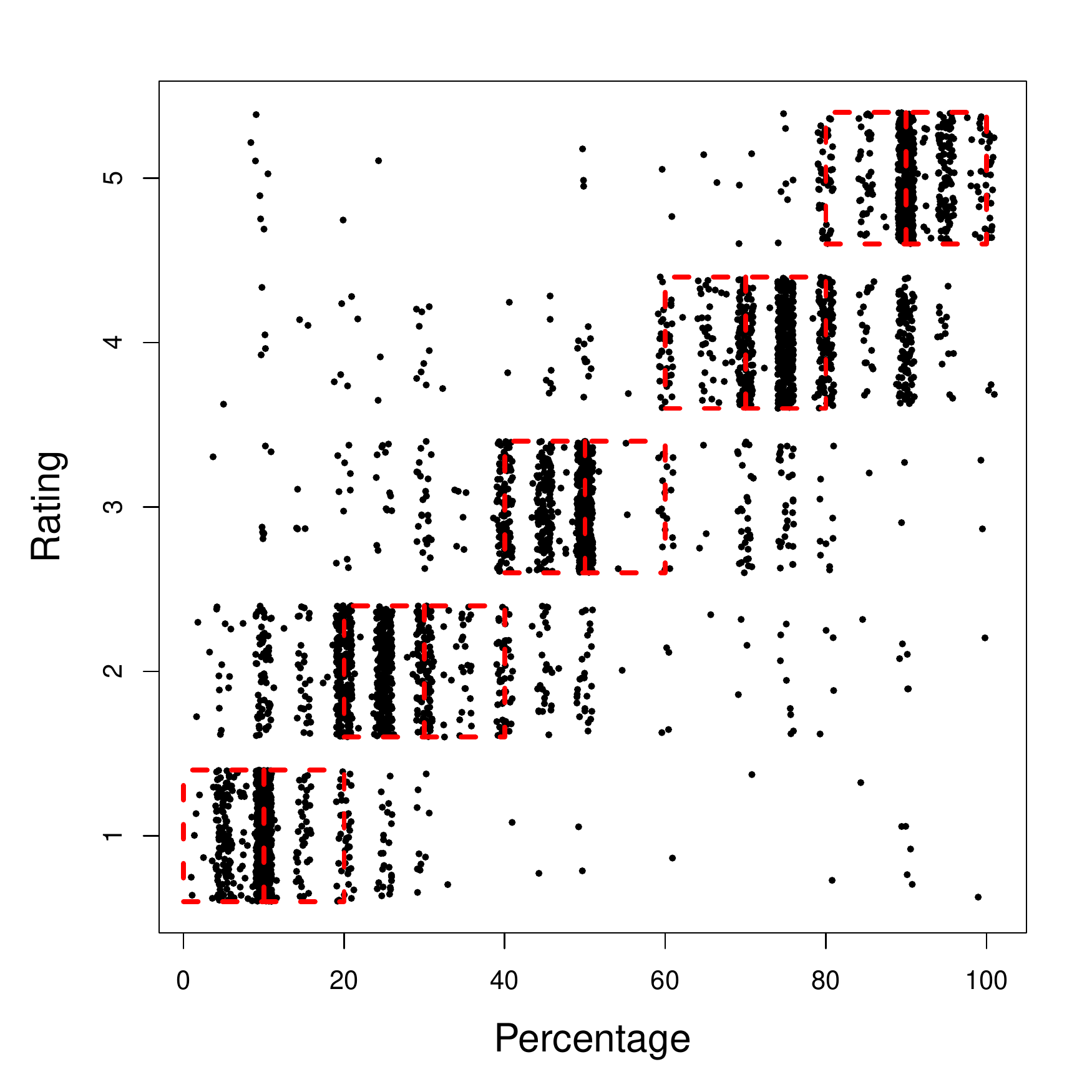}
 \caption{Estimated ratings plotted against estimated percentages for the cylinder questions (jittered). The red boxes shows where correctly transformed percentages should lie.}\label{F:percrate}
\end{figure}

To study the consistency of answering the cylinder items, we compare the estimated percentage with the rating scale value on the 10 cylinder items. Recall that in random order, each true percentage of red filling was not only rated by the respondent on a five point rating scale but they also gave estimated percentage of the amount of red. Figure~\ref{F:percrate} shows a scatterplot of the estimated percentages (horizontal) against the estimated rating vertically. To ensure some spread of individual observations in the plot, the ratings are slightly jittered.  The superimposed red boxes indicate the estimated percentages that should be mapped to the corresponding ratings corresponding to the correct filling. Hence, all observations outside the red boxes represent errors. There could be several sources of such errors. The respondent could misinterpret the direction of the rating scale of the percentage scale, or could making a typing error.  Additionally, the visual stimulus might not be easy to assess for a respondent. Finally, other sources of errors usual in surveys (participant fatigue, response styles, etc.) could also have an effect. 

There are different ways to assess whether these really are mistakes. First, consider errors made in estimating the percentages, thereby only considering the horizontal axis in Figure~\ref{F:percrate}. Per true cylinder filling, the error rate is one minus the proportion of observations that provide the estimated percentage within ten percent of the true percentage of red. These error rates for the five cylinders are $7\%, 22\%, 18\%, 21\%$ and $4\%$ for cylinder fillings of 10\%, 30\%, 50\%, 70\%, and 90\%, respectively. This result indicates that the middle cylinders with true filling of 30\%, 50\%, and 70\% are the hardest to correctly associate with the admissible range of percentages, as can be expected. 

Next, consider the errors of the ratings. Here, it is interesting to look at the errors conditional on the estimated percentages. To do so, we use again the intervals of estimated percentages that were used for the red boxes in  Figure~\ref{F:percrate}. Table~\ref{T:percprop} shows the proportion of responses in the successive intervals $[0, 20], (20,40], \ldots, (80, 100]$. The number of respondents who were able to associate the correct percentage interval with the respective ratings range from 65\% to 81\%. These numbers are not particularly high. However, given that the cylinders are displayed for one second only and that it is not possible to view the cylinder a second time, it may be expected that each respondent will make one mistake. In that case, the hitrate would be 80\%. For rating Categories 1 and 2, the observed hitrates are lower (65\% and 69\% respectively).

\begin{table}
\centering
\begin{tabular}{rccccc}
  \toprule
  & \multicolumn{5}{c}{Rating} \\ 
\cmidrule(lr){2-6}
 Interval & 1 & 2 & 3 & 4 & 5 \\ 
  \midrule
  $[0, 20]$ & 0.65 & 0.31 & 0.02 & 0.01 & 0.01 \\ 
  $(20,40]$ & 0.06 & 0.69 & 0.23 & 0.02 & 0.00 \\ 
  $(40,60]$ & 0.01 & 0.09 & 0.81 & 0.09 & 0.01 \\ 
  $(60,80]$ & 0.00 & 0.02 & 0.09 & 0.81 & 0.07 \\ 
  $(80,100]$ & 0.01 & 0.01 & 0.01 & 0.18 & 0.79 \\ 
   \bottomrule
\end{tabular}
\caption{Proportion of responses in the intervals $[0, 20], (20,40], \ldots, (80, 100]$.} 
\label{T:percprop}
\end{table}

Furthermore, the number of errors made by each respondent can be studied. Define an error as misspecifying a rating by more than one rating category or as misjudging a percentage by more than 15\%. Errors were made by 134 out of 786 respondents (or 17\%). By this definition, a total of 184 mistakes were made. There were 101, 19, 11 and 3 respondents who made 1, 2, 3 or 4 mistakes respectively. Despite these errors, the subsequent results are based on all 786 respondents in the treatment group. The main reason being that the large majority of respondents (95.8\%) made no errors or only one error. 


\subsection{Treatment Differences on Average Rating Scale Distributions}
In this subsection, we study the causal effect of priming the respondents with the cylinder items. In particular, we look at the overall distribution of rating scale usage over the 16 items combined.  

\begin{figure}
\centering
    \includegraphics[width=0.45\textwidth]{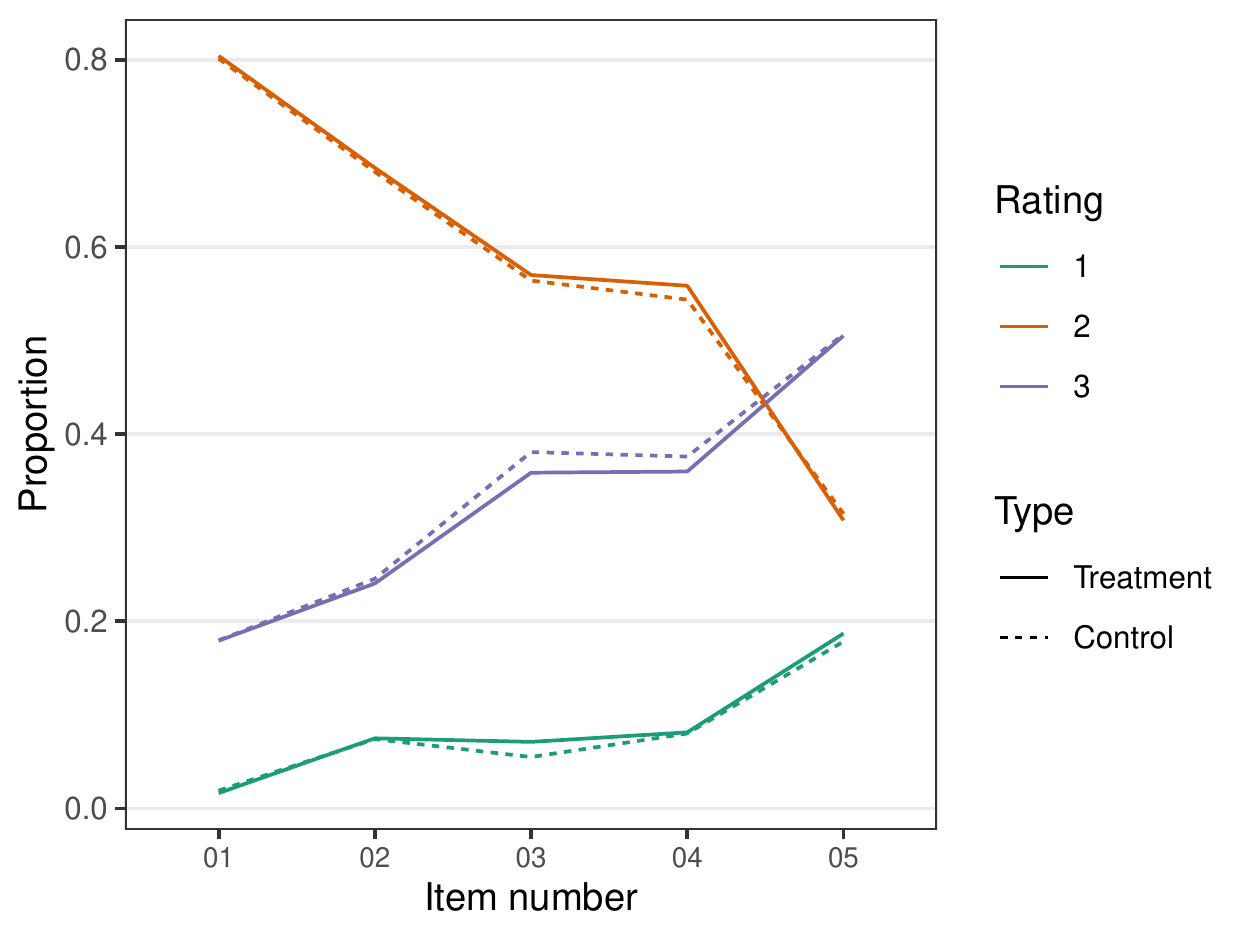}
\caption{The proportion of rating categories used for the five items delivered before the cylinder questions, separately for the treatment and control groups.}
\label{F:props-0105}
\end{figure}

As a first test, we investigate the pre-treatment distribution over the rating scale categories. For this purpose, we consider the five survey items on the economic outlook for the Dutch economy as they appeared in the survey before the experimental condition. As respondents were randomly assigned to the treatment or control groups, the expectation is that there should be no significant differences in distribution over the three-point rating scale items. Figure~\ref{F:props-0105} shows a line per rating scale category over the five items with the proportion usage split by treatment (solid line) and control group (dashed line). As the dashed and solid lines almost overlap, we can conclude that the random assignment has worked well and that there were no pre-treatment differences between the two groups.   

We now switch to the 16 post treatment items and compare the average rating scale usage. As a first step, we do a two-sample Kolmogorov-Smirnov (KS) test to test the null hypothesis that the two distributions of usage of the rating scale categories are equivalent. Comparing the post-treatment scale usage for the 16 items, the null hypothesis is rejected ($D = 0.0617$, asymptotic $p < 0.0001$).  


A more thorough test on differences between the two distributions that is less sensitive for tied data than the KS-test can be obtained through a permutation test. In this test, a statistic derived from the two distributions is compared with the distribution of this statistic under random permutation of the grouping (that is, treatment or control). The test randomly permutes the groupings (i.e. treatment of control) a large number of times (10 000 in this case) and compares the observed difference in category usage between the treatment and control groups with the permutation distribution of the same statistic. If the observed difference of the test statistic falls in the tails of the permutation distribution, it can be concluded that the null hypothesis of equal distributions is not plausible and therefore rejected. The test can be investigated per rating, but also across all ratings by using the sum of squared differences. 

\begin{figure}
\centering
    {\label{F:permglobpost}
        \includegraphics[width=0.5\textwidth]{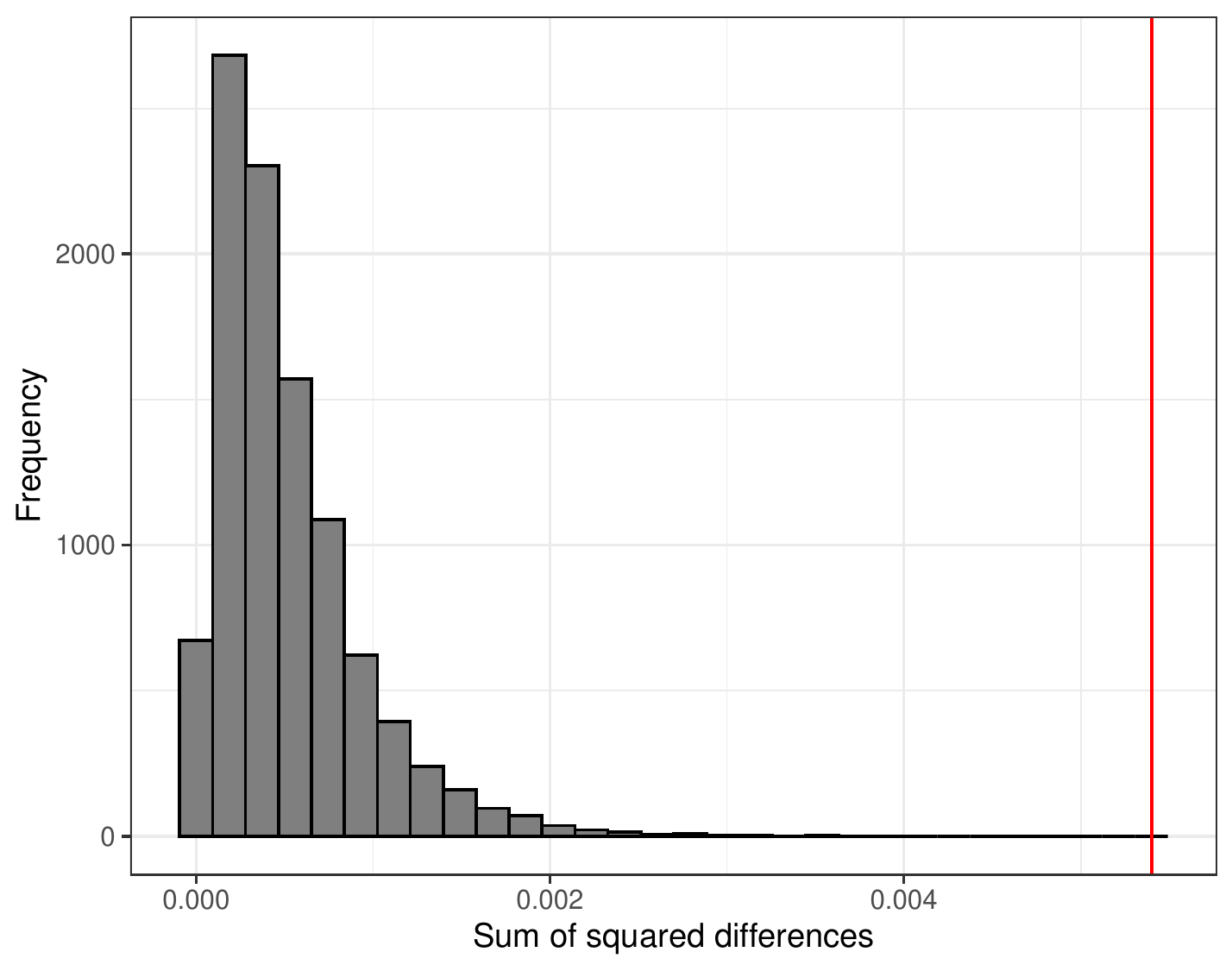}
        } 
        \\
\caption{Global permutation distribution for sum of squared differences in relative frequencies of rating scale use, using 10,000 permutations. The observed value of the test statistic is shown by the red vertical line.}
\label{F:permglob}
\end{figure}

Here, we apply the permutation test on a comparison on average rating scale distribution over the items. As a test statistic, we simply use the sum of squared differences of relative frequencies of the rating scales between the treatment and control groups. This permutation distribution for 10,000 random permutations is shown in Figure~\ref{F:permglob} where the (unpermuted) observed test statistic is indicated by the red line. As all 10 000 test-statistics obtained by the random permutation of the two groups are smaller than the observed test statistic, we can reject the null hypothesis of equal distributions with $p < 0.0001$. 

\begin{figure}
\centering
        \includegraphics[width=0.55\textwidth]{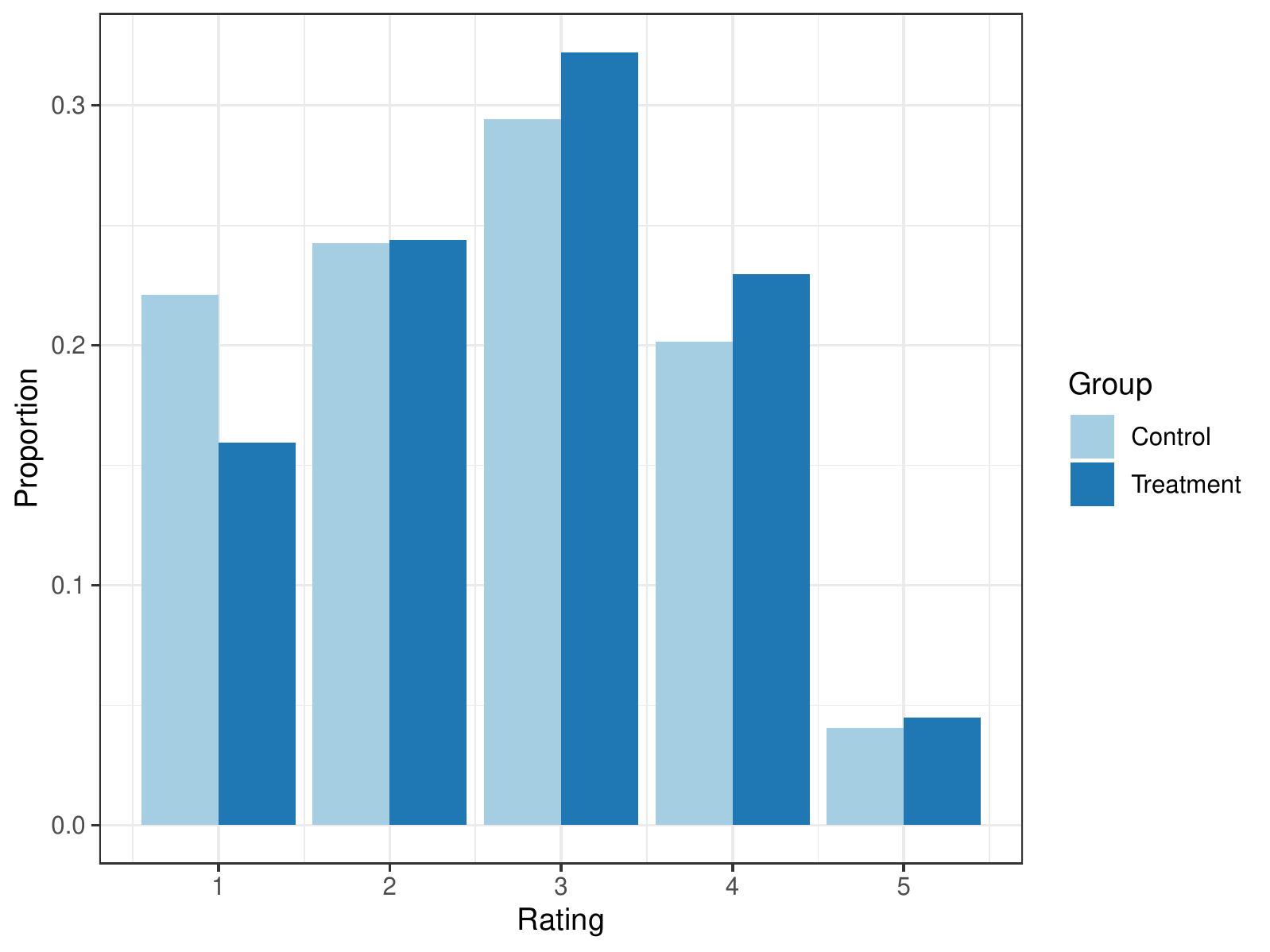}
\caption{Rating scale usage for the treatment and control groups for the 16 survey items following the cylinder intervention.}
\label{F:usage}
\end{figure}

In Figure~\ref{F:usage}, the two observed distributions are shown. It is apparent that administering the cylinder items changes the subsequent rating scale usage of the respondents. Specifically, the treatment group uses fewer 1's (19.6\% versus 27.5\%),  more 3's (32.5\% versus 28.7\%) and more 4's (19.6\% versus 16.3\%). In conclusion, there is a significant difference in rating scale use after the cylinder questions have been administered. The cylinder items seem to make respondents more aware of the spread of the rating scale. Combined with the evidence from Section~\ref{S:consistency} it can be argued that the anchoring through the cylinder items seems to make respondents pay more attention to the meaning of rating scale categories within and across respondents. 


%

\subsection{Treatment Differences on per Item Rating Scale Distributions}
\begin{figure}
\centering
    \includegraphics[width=0.85\textwidth]{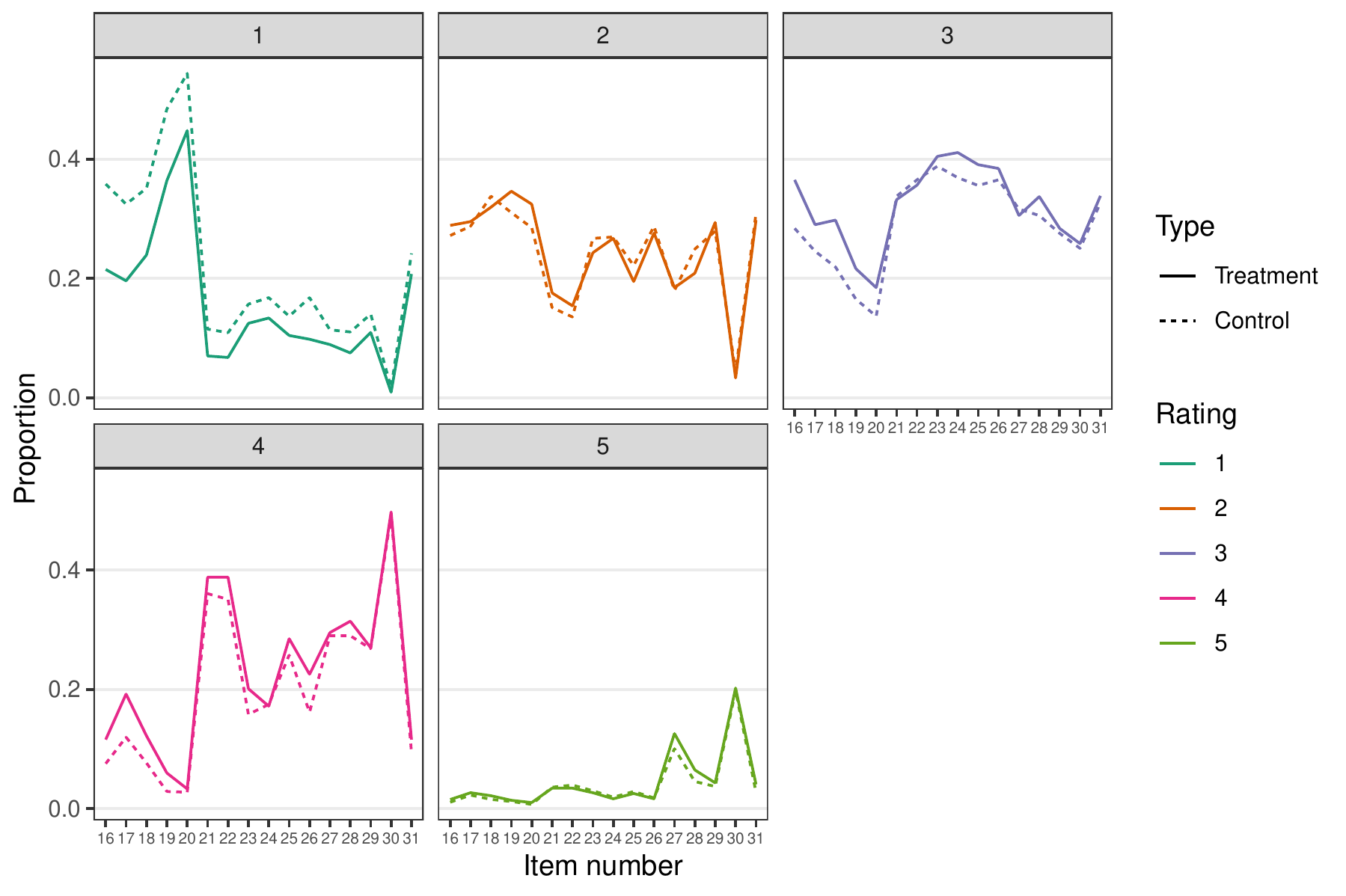}
\caption{The proportion of rating categories used for the 20 items delivered after the cylinder questions, separately for the treatment and control groups. Each panel shows a different rating category.}
\label{F:props-1635}
\end{figure}

Having established in the previous section that there is an \textit{average} effect of the cylinder items on the distribution over the categories of the rating scale, we now turn to item specific effects. Table~\ref{T:freqs-control}  and Figure~\ref{F:props-1635} give the relative frequencies (proportions) per rating and item for the treatment and control groups. In the previous section, we concluded that the differences mainly lie in the usage of Categories 1, 3, and 4. In particular,  Figure~\ref{F:props-1635} shows 
that for Category 1, there is a much smaller proportion for almost all items, with the strongest effects for the first 14 items after the treatment. In addition, Categories 3 and 4 are used more often in the treatment group, again with the strongest effects for the first items after the treatment. In conclusion, the anchoring effect of the cylinder items seems strongest for the first (say 10--15) items, after which the effect becomes hard to detect. Morevoer, Category 1 is used less often in the treatment group and Categories 3 and 4 more.

\begin{table}
\centering
\begin{tabular}{lccccc}
\toprule
 & \multicolumn{5}{c}{Rating Treatment} \\ 
\cmidrule(lr){2-6}
Item & 1 & 2 & 3 & 4 & 5 \\ 
\midrule
\multicolumn{6}{l}{\textit{Innovation}} \\
V16 & $0.22$ & $0.29$ & $0.37$ & $0.12$ & $0.02$ \\ 
V23 & $0.12$ & $0.24$ & $0.40$ & $0.20$ & $0.03$ \\ 
V24 & $0.13$ & $0.27$ & $0.41$ & $0.17$ & $0.02$ \\ 
V25 & $0.10$ & $0.19$ & $0.39$ & $0.28$ & $0.03$ \\ \\
\multicolumn{5}{l}{\textit{Materialism}} \\
V17 & $0.20$ & $0.30$ & $0.29$ & $0.19$ & $0.03$ \\ 
V18 & $0.24$ & $0.32$ & $0.30$ & $0.12$ & $0.02$ \\ 
V19 & $0.36$ & $0.35$ & $0.22$ & $0.06$ & $0.01$ \\ 
V20 & $0.45$ & $0.32$ & $0.18$ & $0.03$ & $0.01$ \\ 
V21 & $0.07$ & $0.18$ & $0.33$ & $0.39$ & $0.03$ \\ 
V22 & $0.07$ & $0.15$ & $0.36$ & $0.39$ & $0.03$ \\ 
V26 & $0.10$ & $0.28$ & $0.38$ & $0.23$ & $0.02$ \\ 
V27 & $0.09$ & $0.18$ & $0.31$ & $0.30$ & $0.13$ \\ 
V28 & $0.08$ & $0.21$ & $0.34$ & $0.31$ & $0.06$ \\ 
V29 & $0.11$ & $0.29$ & $0.28$ & $0.27$ & $0.04$ \\ 
V30 & $0.01$ & $0.03$ & $0.26$ & $0.50$ & $0.20$ \\ 
V31 & $0.21$ & $0.30$ & $0.34$ & $0.12$ & $0.04$ \\ 
\\
Total & $0.16$ & $0.24$ & $0.32$ & $0.23$ & $0.04$ \\ 
\bottomrule
\end{tabular}
\quad
\begin{tabular}{lccccc}
\toprule
 & \multicolumn{5}{c}{Rating Control} \\ 
\cmidrule(lr){2-6}
Item & 1 & 2 & 3 & 4 & 5 \\ 
\midrule
\multicolumn{5}{l}{\textit{Innovation}} \\
V16 & $0.36$ & $0.27$ & $0.28$ & $0.08$ & $0.01$ \\ 
V23 & $0.16$ & $0.27$ & $0.39$ & $0.16$ & $0.03$ \\ 
V24 & $0.17$ & $0.27$ & $0.37$ & $0.17$ & $0.02$ \\ 
V25 & $0.14$ & $0.22$ & $0.36$ & $0.26$ & $0.03$ \\ \\
\multicolumn{6}{l}{\textit{Materialism}} \\
V17 & $0.32$ & $0.29$ & $0.25$ & $0.12$ & $0.02$ \\ 
V18 & $0.35$ & $0.34$ & $0.22$ & $0.08$ & $0.02$ \\ 
V19 & $0.48$ & $0.31$ & $0.17$ & $0.03$ & $0.01$ \\ 
V20 & $0.54$ & $0.29$ & $0.14$ & $0.03$ & $0.01$ \\ 
V21 & $0.11$ & $0.15$ & $0.34$ & $0.36$ & $0.04$ \\ 
V22 & $0.11$ & $0.14$ & $0.37$ & $0.35$ & $0.04$ \\ 
V26 & $0.17$ & $0.29$ & $0.37$ & $0.16$ & $0.02$ \\ 
V27 & $0.11$ & $0.18$ & $0.32$ & $0.29$ & $0.10$ \\ 
V28 & $0.11$ & $0.25$ & $0.31$ & $0.29$ & $0.05$ \\ 
V29 & $0.14$ & $0.28$ & $0.28$ & $0.27$ & $0.04$ \\ 
V30 & $0.02$ & $0.04$ & $0.25$ & $0.49$ & $0.20$ \\ 
V31 & $0.24$ & $0.31$ & $0.33$ & $0.09$ & $0.03$ \\ 
\\
Total & $0.22$ & $0.24$ & $0.29$ & $0.20$ & $0.04$ \\ 
\bottomrule
\end{tabular}
\caption{Proportion of rating scale use per item for the treatment (left) and control groups (right), for 16 items administered after the cylinder question. The marginal proportions are shown in the final row.}
\label{T:freqs-control}
\end{table}

\subsection{Differential Item Effects}

In addition to the average effect of the priming on rating scale usage, there can be a differential effect for the items. To explore this, we perform a correspondence analysis (CA) on the table of items by rating categories for the two groups. That is a frequency table with the rating categories for the two groups in the columns and the items in the rows. It is well known that the CA of this matrix removes the main and interaction effects of category usage and being in the treatment or control group \citep[see][]{van1989combined}. The CA will therefore only show the remaining effects between items and rating scale category of treatment and control groups. 

\begin{figure}
\centering
    \includegraphics[width=0.55\textwidth]{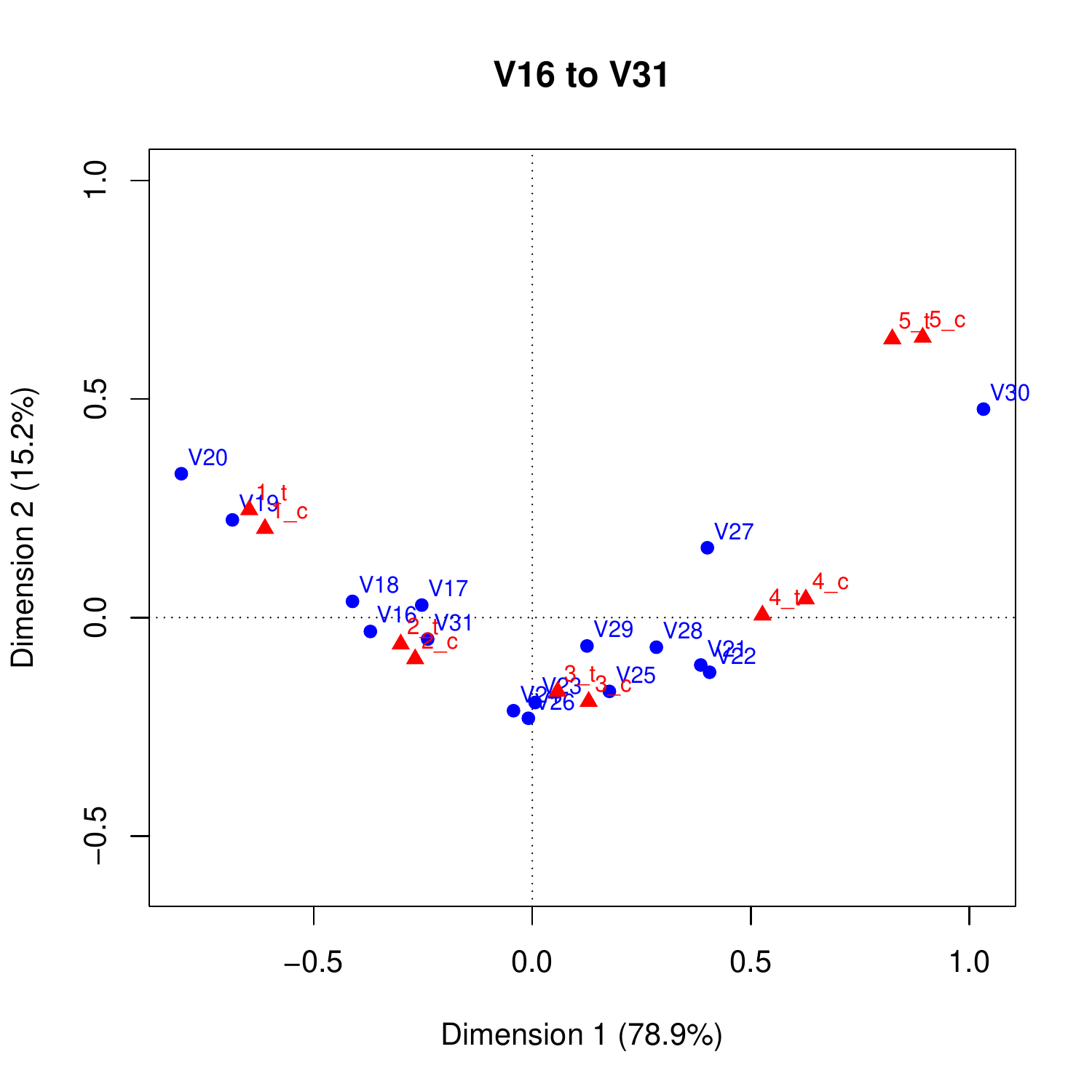}
\caption{Correspondence analysis on the rating scale usage of Items 16 to 31. The frequencies for the treatment (t) and control (c) groups are calculated separately. These are then concatenated row-wise, and subjected to correspondence analysis.}
\label{F:ca-combined-1631}
\end{figure}

Figure~\ref{F:ca-combined-1631} shows the results of the CA. First we observe that the first dimension explains almost 80 percent of the inertia and is about five times stronger than Dimension 2.
Although the points of categories for treatment and control are close, those of the treatment are slightly, but consistently, to the left of  those of the points corresponding to the categories for the control.  Apart from the main and interaction effects, items V19 and V20 have more scores on Category 1 and fewer scores on Categories 4 and 5. This effect is slightly stronger for the Treatment group. For V30 the opposite is true: it has more scores in Categories 5 and 4 and fewer in Category 1, and the effect is slightly stronger for the Control group than the Treatment group.

\section{Conclusions and Discussion}
It is evident that the simple cylinder questions has a priming effect on the respondents which greatly affects their rating scale usage. Our experiment shows that respondents answering on a five point rating scale, respondents in the primed group use Category 1 less and Categories 3 and 4 more than those who were not primed. This suggests that the treatment reduces extreme responding, a main response style, and may motivate respondents to pay closer attention to items.

The section on consistent rating scale usage provides evidence that about 17\% of the respondents make errors in their ratings, but only 4\% make more than one error. It is possible that the visual cues by the cylinders are misinterpreted by these respondent. In this analysis, we have used all observations and not made a selection of the participants based on these inconsistencies in the treatment group. It is our expectation that if those participants would have been removed, the effect of the visual priming could be stronger.

We conclude that priming respondents by a visual anchor such as the filling of a cylinder has an effect and can improve the quality of rating scale data that may improve a numerical comparison within and between respondents. We call for additional experimentation with similar anchors to study the strength of the effect and its applicability in a wider range of surveys that use rating scales.

\bibliography{cylinders}

\begin{thebibliography}{}

\bibitem[Blasius and Thiessen, 2012]{blasius2012}
Blasius, J. and Thiessen, V. (2012).
\newblock {\em Assessing the quality of survey data}.
\newblock Sage.

\bibitem[Cunningham et~al., 1977]{cunningham1977}
Cunningham, W.~H., Cunningham, I.~C., and Green, R.~T. (1977).
\newblock The ipsative process to reduce response set bias.
\newblock {\em Public Opinion Quarterly}, 41(3):379--384.

\bibitem[Richins and Dawson, 1992]{richins1992}
Richins, M.~L. and Dawson, S. (1992).
\newblock A consumer values orientation for materialism and its measurement:
  Scale development and validation.
\newblock {\em Journal of Consumer Research}, 19(3):303--316.

\bibitem[Rudnev, 2021]{rudnev2021}
Rudnev, M. (2021).
\newblock Caveats of non-ipsatization of basic values: A review of issues and a
  simulation study.
\newblock {\em Journal of Research in Personality}, 93:104118.

\bibitem[Schoonees et~al., 2015]{schoonees2015constrained}
Schoonees, P.~C., Van~de Velden, M., and Groenen, P. J.~F. (2015).
\newblock Constrained dual scaling for detecting response styles in categorical
  data.
\newblock {\em Psychometrika}, 80(4):968--994.

\bibitem[Steenkamp et~al., 1999]{steenkamp1999}
Steenkamp, J.-B.~E., Ter~Hofstede, F., and Wedel, M. (1999).
\newblock A cross-national investigation into the individual and national
  cultural antecedents of consumer innovativeness.
\newblock {\em Journal of Marketing}, 63(2):55--69.

\bibitem[Van~der Heijden et~al., 1989]{van1989combined}
Van~der Heijden, P. G.~M., De~Falguerolles, A., and De~Leeuw, J. (1989).
\newblock A combined approach to contingency table analysis using
  correspondence analysis and loglinear analysis.
\newblock {\em Journal of the Royal Statistical Society: Series C (Applied
  Statistics)}, 38(2):249--273.

\bibitem[Van~Herk et~al., 2004]{herk2004}
Van~Herk, H., Poortinga, Y.~H., and Verhallen, T.~M. (2004).
\newblock Response styles in rating scales: Evidence of method bias in data
  from six eu countries.
\newblock {\em Journal of Cross-Cultural Psychology}, 35(3):346--360.

\bibitem[Van~Rosmalen et~al., 2010]{vanRosmalen2010identifying}
Van~Rosmalen, J., Van~Herk, H., and Groenen, P. J.~F. (2010).
\newblock Identifying response styles: A latent-class bilinear multinomial
  logit model.
\newblock {\em Journal of Marketing Research}, 47(1):157--172.

\end{thebibliography}
\bibliographystyle{apalike}

\end{document}